# AI in Money Matters

Nadine Sandjo Tchatchoua

Roskilde University, nadinet@ruc.dk

Richard Harper

LANCASTER UNIVERSITY, R.HARPER@LANCASTER.AC.UK

**Abstract**

In November 2022, Europe and the world by and large were stunned by the birth of a new large language model : ChatGPT. Ever since then, both academic and populist discussions have taken place in various public spheres such as LinkedIn and X(formerly known as Twitter) with the view to both understand the tool and its benefits for the society. The views of real actors in professional spaces, especially in regulated industries such as finance and law have been largely missing. We aim to begin to close this gap by presenting results from an empirical investigation conducted through interviews with professional actors in the Fintech industry. The paper asks the question, how and to what extent are large language models in general and ChatGPT in particular being adopted and used in the Fintech industry? The results show that while the fintech experts we spoke with see a potential in using large language models in the future, a lot of questions marks remain concerning how they are policed and therefore might be adopted in a regulated industry such as Fintech. This paper aims to add to the existing academic discussing around large language models, with a contribution to our understanding of professional viewpoints.

CCS Concepts: • **Large Language Models → ChatGPT**

**KEYWORDS** • Large Language Models • ChatGPT • Fintech • Artificial Intelligence .

**1 Introduction**

Since the term 'artificial intelligence' came into common parlance (arguably in 1955 when John McCarthy held a workshop with that name), considerable research work has been undertaken to examine the nature of AI and more specifically, its relationship to human activity. This can be dated at very least to notions of the 'Turing test'; to early examples of the way in which human beings used, or even entrusted, computer simulations and, in turn, how human behaviour might be modelled in computer systems (Weizenbaum, 1966; Colby, 1981; Bassett, 2019, 2021; Natale, 2019) The so-called '5th generation' AI, despite much of the hype, has led to practical innovations with respect to computer supported decision-making. In the present day, we have arrived at a point where - again with the hyped versions - computer applications may have a decision-making capacity with or without human intervention (see for instance Morison and Mcinerny, 2024; Feuerriegel et al, 2024; Mosqueira-Reyet al, 2023; Zanzotto, 2023).

The advent of Large Language Models (LLMs) such as CHATGPT, Bard (or Gemini), and so on -all examples of so-called 'generative AI' - has reinvigorated the debate around AI, its powers, functions and possibilities. Some of this has to do with overblown claims about the 'singularity', arguments about the philosophical grounds for believing, at least, that machines might (at long last) surpass the intelligence of human beings (Should this happen, it will be predicated on the development of Artificial General Intelligence (AGI) (. See e.g. Kurtzweil, 2005; Walsh, 2017; Braga and Logan, 2019: Hoffman, 2022 for a summary of the debate).Similarly, we have witnessed dramatic claims about the political and economic ramifications of the new technology (e.g. Luitse and Denkena, 2021: Bender and Gebru, 2021). More cautious voices have warned against what in their view is the same species of hyperbole that accompanied earlier generations of AI. These latter assessments have to do with the limitations of Generative AI with respect to 'ground truth'. Generative AI here refers to the latest iterations of AI technology, so called because they generate text and images from a wide assemblage of data in a 'natural' form. The data in question is sometimes assumed to have an objective form, hence the concept of 'ground truth' (but see e.g. Blackwell, 2017).

Further doubts about the kinds of application we might reasonably expect to devolve from LLMs exist, with environmental concerns, with concerns about how human beings can and will interact with such models, and with the recognition that there may be some natural limit to their growth. Partly because of these concerns, there has been a developing interest in the degree to which LLMs can be used to train smaller, and domain specific models. Nevertheless, and this is the subject of our paper, many issues need to be examined and understood if LLMs are to prove useful and usable in the Fintech industry specifically. Our research questions, then, are:

'What factors influence the adoption of LLMs in the Fintech industry?' 'How have companies in Fintech reacted to recent development in LLMs?'

## 2 State of the Art

### 2.1 The challenges of generative AI: Ground Truth

The usability of generative AI, , comes down to several factors of which the reliability of outputs is one of the most critical. Early attempts to generate human-like 'voices' through deep learning foundered on their responsiveness to certain kinds of input, in this case input directly from human beings. The best-known examples are Tay and Zo. Tay, which was intended to simulate the voice of a teenage girl quickly began to spout various kinds of racist offensiveness, perhaps because of a concerted X (previously Twitter) effort by right wing trolls (see Wolf et al, 2017; Fuchs, 2018). Zo, which was an attempt to correct this by producing more 'politically correct' responses, equally generated some controversy by its refusal to engage with, or respond in any sensible way, to inputs such as 'They have good falafel in the Middle east (see https://qz.com/1340990/microsofts-politically-correct-chat-bot-is-even-worse-than-its-racist-one). Outputs in both cases were a direct product of the specific learning datasets used.

Other kinds of input are more indirect and can be thought of as documentary, relying for the most part on publicly available (but nevertheless selected) data. These large language models (LLMs) have become possible for several interconnected reasons. These include the availability of vast amounts of data, mainly from the Internet but also, from arrays of sensors in a whole variety of locations. This affects the speed with which data can now be examined, processed and organised. Some, at least, of the analytical techniques in question are not in fact that new. Most are well-known methods for reducing complex data to recognisable patterns and from that extracting 'types' of data. They include Bayesian statistics, logistic regression analysis, decision trees, random forests, support vector machines, and so on (Blackwell, 2021; Domingos, 2017).

Others are. So-called transfer learning is a significant development, because it opens the possibility of a more generalised form of AI (Russell and Norvig, 2017; Russell, 2019). An LLM learns associations from scratch during its training phase—over billions of training runs, its attention network slowly encodes the structure of the language it sees as numbers (called "weights") within its neural network. Text is split into tokens, which are words like "love" or "are", affixes, like "dis" or "ised", and punctuation, like "?" .Transfer learning (Vaswani et al, 2017) is a technique which relies on a method which allows any given sequence to be re-ordered by weighting the importance of elements of the input in an N-dimensional space. The result is what is now referred to as a foundation model (Bommasani et al, 2022).

Regardless, the problem of 'ground truth' remains, and can be thought of as essentially a problem of reliability both in respect to whether outputs are factually correct, or otherwise acceptable ethically and politically, or not. Various commentators have pointed to problematic elements. Hence, Guidotti et al:

"This enormous amount of data may contain human biases and prejudices. Thus, decision models learned on them may inherit such biases, possibly leading to unfair and wrong decisions." (Guidotti et al, 2018:1)

Others, such as Blackwell (2017, 2021, 2023); Collins, (2018), and Schneiderman (2022) have examined the relationship between the apparent objectivity presumed in 'ground truth' and the distinct ways in which human subjectivity (or intersubjectivity) play out in cultural contexts. While we do not have space to examine these arguments in detail in this paper, they do point clearly to one highly problematic feature of LLMs, which is their reliability. References to so-called, 'hallucinations' have become common along with attempts to mitigate them (see e.g. Perkovic et al, 2024).

A second, related, issue, has to do with usability and usefulness. This ramifies in various ways. Harper and Randall (2013) have pointed to various issues surrounding the use of generative AI in the real world. They suggest that how users decide on appropriate inputs and make sense of outputs and further find ways to describe data outputs in such a way that they are usable are legitimate concerns for Computer Supported Cooperative Work (CSCW). As they say, "… we might take our lead from Garfinkel (1967) by showing how accounting for decisions which go into feeding data into systems to produce a 'ground truth' are accountable matters, matters worthy of careful investigation, just as are the ways that objective functions are accountable too. In this view, explanations are not generalised, as seems to be thought by the AI community, but designed to be appropriate for some tasks at hand – they lead to subsequent action. The question is what action, why and with what consequences – from inputs through to outputs."

In turn, such a statement has ramifications in relation to the division of labour and who has responsibility for deciding on the AI usage, its purposes and its limitations. Not least, in some contexts there will be important regulatory aspects to be considered. GDPR regulations and the EU AI policy (see Justo-Hanani, 2022)currently require that AI systems should empower human beings, allowing them to make informed decisions through human-in-the-loop approaches, that AI systems need to be safe, resilient and secure, that AI systems should ensure privacy and proper governance, that they should be transparent and decisions traceable/ explainable, that they should be free from bias and should not discriminate, that they should entail sustainable benefits to the wider human population, and that mechanisms should be put in place to ensure accountability for AI systems and their outcomes.

Further to this, exactly how machine outputs are to be represented to user populations is an entirely non-trivial matter. If AI of whatever kind is to be made 'explainable', then we need to investigate what properties human beings might or might not have such that they will be capable of understanding not only the outputs but the reasons for them. As Harper and Randall put it, "… obvious questions arise as to what person, when, in what circumstances, and so on". They cite some CSCW studies which have demonstrated some practical uses (see Ontika et al, 2022; Park et al, 2019; Osman et al, 2021; Ploug and Holm, 2020).

Nevertheless, we would argue that the organisational contexts -such as the Fintech industry -in which generative AI technologies are going to be interpreted, how decisions will be implemented, how they are made compliant with organisational, ethical and institutional constraints, by whom, and why, are largely under-researched. Workers with different occupational roles will have specific aims associated with the task and hence inputs and outputs will necessarily have to be made relevant to those requirements. This, in turn, raises questions about appropriate ways of visualising machine inputs and outputs. Not least, the uses of generative AI will resonate with the social distribution of expertise and to some degree the expertise in question will have to do with 'prompt' engineering, the design of questions to elicit appropriate responses from an LLM (See Chen et al, 2023 for a review). This will matter both in terms of decisions about what kinds of input are allowable and/or acceptable and what kinds of output are relevant, understandable and trustable. There is evidently a scope for more investigation into work domains which might enable us to make better sense of what needs to be explained and to whom. We develop this below.

A third problem is that of scale. Some of the problems associated with LLMs include the possibility of a natural limit. There is a view that LLMs are rapidly approaching a 'maximum' size, for a range of reasons. These include the need for new hardware, including radical chip design; dealing with the huge energy demands associated with their use, demands which exceed, for instance those of Blockchain applications (see Sedlmeir et al, 2020); the demand for accuracy, and some limit to the size of the models themselves. Model sizes have grown 10 times in 6 years (Google). GPT-4 has 1 trillion parameters, and cost $ 100 million +. Computing power needed is estimated to double every 6 years (see e.g. https://spectrum.ieee.org/ai-energy-consumption). Estimates of cost are that Microsoft may invest as much as $10 billion in OpenAI development. (see

https://www.euronews.com/next/2023/01/23/chatgpt-microsoft-invests-billions-more-in-openai-as-tech-race-with-google-heats-up-around).

Google plans to use LLMs on 6 services with 2 billion users. These limitations mean that attention is now being paid to the prospect of using LLMs to train domain specific models, models which 'piggyback' on LLMs will require a considerable amount of power too. Goldman Sachs, a well-known financial institution for instance, estimates that 44% of its legal tasks could be done by AI. In fact, Legal assistants like "Cocounsel" (a customised version of GPT4). are already being used (see Callister, 2023, for a discussion)

**2.2 The uses of generative AI.**

If the history of new technology teaches us anything, it is that the uses of the technology in question tend to emerge slowly. Yet, we have few indications of what the uses of generative AI will, in practice, turn out to be. A recent survey by the Boston Consulting Group (https://www.bcg.com/publications/2024/from-potential-to-profit-with-genai) suggested that most executives thought it would take at least two years (from now) "to move beyond the hype". The American census Bureau reports that only 5.4% of US businesses currently use any form of AI at all (https://www.census.gov/hfp/btos/downloads/CES-WP-24-16.pdf). What little information we have suggests that office work can be made more efficient. Noy and Zhang (SSRN, 2023) in a study of 44 office workers, demonstrate how, when ChatGPT is put in the hands of office workers, it appeared to speed up the completion of tasks (assigned tasks were completed in 17 minutes, compared to an average 27 minutes), the quality of work was seen to have improved, and participants reported higher satisfaction with their work. There have been a few studies of workers anticipating possible uses (or non-uses) of generative AI. One such study is that of Woodruff et al (2024), who interviewed knowledge workers from seven different industries in a series of workshops. Several narratives emerged from their participants, of which the dominant one was a generalized belief that generative AI would be mainly used for the substitution of menial tasks. There appeared to be agreement that decision-making by such tools was some way off and, in any case, not to be desired. We have built on this to examine in more detail attitudes and policies in one sector.

ChatGPT itself, when asked what occupational tasks it might perform, includes Customer Support, by providing instant responses to customer queries. It added troubleshooting common issues and helping with products or services. Content Creation was also included in this answer, through generating articles, blog posts, product descriptions, social media content, and marketing materials. Translating text from one language to another was also mentioned, as well as tutoring and education, data analysis and reporting, software development, conversational support in therapeutic environments, research assistance, paralegal enquiries, market research, and clerical and admin support. Felton et al at Princeton University, in a survey of 700 occupations, suggest the professions most likely to be affected by ChatGPT are those which have routine, repetitive, evidence-based and scriptable elements. Various studies have established that, in principle, LLMs might be used in such instances as generating and understanding code (e.g. Nam et al, 2024; Saha et al, 2024), as adjuncts to recommender systems (see Lubos et al, 2024), in a variety of medical contexts, e.g. for text annotation (See Goel et al, 2023), as a diagnostic assistant (see Wu et al, 2023), for training and for patient interaction (see Thirunavukarasu et al, 2023 for a general review). Having said all this, it remains the case that we have little or no direct evidence with which to assess the reality of these claims. Nothing that looks like real-world evaluation in use has been published, to our knowledge, and very little which solicits expert views on how LLMs will be used in practice, and what the potential pitfalls, obstacles and other restrictions might be.

**2.3 AI and Ethics**

Inevitably, LLMs will come to be deployed in areas which are, for different reasons, politically and ethically sensitive. We have already seen controversy erupt over AI systems deployed in such contexts and the unforeseen consequences which can result. Even before the advent of Machine Learning (ML) techniques in professional contexts, concern had been expressed about AI being used. In the UK, it has been used as a risk assessment method in childcare (Jackson, 2018). In the US, algorithms which assess the likelihood of criminal behaviour are being used as a technique in so-called 'predictive policing'. That is, evaluative procedures are brought to bear, with different conclusions. In the case of predictive policing, police forces understandably use the algorithm to predict crime better and thus prevent it. Criminologists, however, see different consequences. Here, much of the issue has to do with policy decisions that are made after the outcomes of ML processes are known. O Donnell (2019) has argued that the training datasets used in predictive policing embody racist assumptions (see also Asaro, 2019, Ruha Benjamin 2019). Data that is input might include criminal histories, geographical information, family and friendship histories, and so on. The Chicago Police department's 'Strategic Subject List' ranks individuals on the likelihood that they might commit crimes in the future. There is a background problem here which has to do with false negatives and false positives (see Dougherty, 2015 for an egregious example). O'Donnell cites one case where an individual was interviewed by the police, apparently (quoting the Chicago Tribune) because the algorithm had identified them on the basis that 'a childhood friend with whom they had once been arrested on a marijuana charge was fatally shot last year in Austin.' Criminologists will point out that young black men are incarcerated at a rate 5 times that of white men, and that about a third can expect to go to prison at some point in their lives. Policing policy is in the main either person-based or location-based. Either individually or taken together, a score is generated. Of course, the training data will incorporate previous involvement with the police (for instance, number of times stopped by the police). O'Donnell reports that black Americans are twice as likely to be stopped in a vehicle by the police even after controlling for such things as possession of illegal material. As she suggests, "Disparate policing may also stem from white civilians' implicit biases, which cause them to conceive of people of colour as "more dangerous" in some way and cause them to call the police more frequently to address people of colour than they would for similar behaviour of white people". Further biases may exist because of the under-reporting of so-called ''white collar' crime. There is no reason to assume that the problem of biases goes away as LLMs are deployed.

There are, in sum, good reasons to regard training datasets as flawed. Moreover, the variables and their relative weighting are not known to the users of the system who nevertheless implement policy on the back of the system outputs. It requires no great leap of imagination to see how such outputs might influence patrolling, stop and search policy, and so on. Similar issues have arisen with child protection policy. Gillingham (2009; 2015) has argued that outputs are not always used as they might be. In discussing the use of an algorithm in New Zealand, he shows that 'substantiation'- the making of a decision based on proof or evidence – involves similar ambiguities and variations in practice. If 'substantiation' is used as the 'ground truth', then one must know that data about it can be collected in a variety of ways. This, in turn, give rise to ambiguities or hidden biases in datasets. A classic study of pickpocket behaviour in Beijing (Gu et al, 2019) illustrates the general problem with respect to false positives and false negatives.

Regulation adds a further complication. Even a cursory glance shows just how difficult implementation of GDPR regulations will be. The main points being:

---

1. AI systems should empower human beings, allowing them to make informed decisions and providing oversight mechanisms through human-in-the-loop approaches.
2. that AI systems need to be safe, resilient and secure.
3. that AI systems should ensure privacy and proper governance.
4. that AI systems should be transparent and decisions traceable/ explainable.
5. that AI systems should be free from bias and should not discriminate, intentionally or otherwise, against any group.
6. that AI systems should entail sustainable benefits to the wider human population.
7. that mechanisms should be put in place to ensure responsibility and accountability for AI systems and their outcomes.

---

A consummation devoutly to be wished, one might think. Applications using LLMs are a long way from satisfying these requirements and, indeed, now it is far from clear how they can. Concerns with privacy and security remain prevalent. One such issue has to do with the possibility of ML hacking or more generally, system security. How are such systems to be secured? So-called adversarial machine learning may pose problems which are quite different from the more traditional threats, and it has been argued that organisations yet are ill-prepared for dealing with them. As Kumar et al put it, adversarial ML attacks constitute an unknown unknown (Shankar et al, 2020). Attacks might result in the theft of models, theft of training data, manipulation of training data, reprogramming of systems, hacking of physical equipment (e.g. driverless cars) and so on.

In addition, with supervised learning, the system must be 'dosed' with existing data. The data does not come from nowhere and, as has been pointed out by, amongst others (see e.g. Casilli, 2019); and so on, an army of low paid workers, typically based somewhere in Southeast Asia are inputting large amounts of data on a regular basis. One important corollary, of course, is that, in the nature of the 'learning' process, this is not one-time work. As Blackwell puts it:

"… *thousands of people are paid pennies to create a 'ground truth' by providing labels for large data sets of training examples…. In this case, the 'objective function' is no more or less than a comparison of the trained model to previous answers given by the [humans]. If the artificially intelligent computer appears to have duplicated human performance, in the terms anticipated by the celebrated Turing Test, the reason for this achievement is quite plain – the performance appears human because it is human! … The artificial intelligence industry is a subjectivity factory, appropriating human judgments, replaying them through machines, and then claiming epistemological authority by calling it logically 'objective'*".

There are several other issues to contend with. Most of us would agree with the proposition that AI systems should be reliable, safe, and trustworthy; should be relatively free from bias and not promote or reinforce existing structural inequalities; and should empower people by supporting their skills and expertise and should be fit-for-purpose. To a degree, these are the kinds of question that exercise those interested in 'explainable' or 'human in the loop' AI (see e.g. Xu, 2019; Zanzotti, 2023: ). How machine outputs are rendered accountable is a vibrant concern. Guidotti et al (2018) (cited in Harper and Randall, 2024) suggest that many systems designed to support decisions typically hide their internal logic. That is, they constitute yet another 'black box' technology. As they say, "The applications in which black box decision systems can be used are various, and each approach is typically developed to provide a solution for a specific problem and, as a consequence, delineating explicitly or implicitly its own definition of interpretability and explanation." (Guidotti et al, 2018: 1) They go on: "What does it mean that a model is interpretable or transparent? What is an explanation? When a model or an explanation is comprehensible? Which is the best way to provide an explanation, and which kind of model is more interpretable? Which are the problems requiring interpretable models/predictions? What kind of decision data are affected? Which type of data records is more comprehensible? How much are we willing to lose in prediction accuracy to gain any form of interpretability?" (Guidotti et al, 2018: 3)

Most of the arguments we have rehearsed above are academic in nature. There remains, for the reasons rehearsed, a shortage of real- world data concerning how generative AI is or will be used in context specific areas of our society by and large. In what follows we examine views about the possibilities inherent in generative AI, and the challenges that go with it, in a specific domain: The Fintech Industry. Our purpose in so doing is to examine the possibilities and the related challenges from the point of view of actors in this industry. To this end, the first author undertook in depth interviews with both figures with existing experience of using LLMs and others who expect to do so.

Computer Supported Cooperative Work (CSCW) as a field of research insists on real-world contextual studies as a corrective to over-generalisation (see Ackerman et al, 2000, Koch et al, 2015). The hype around LLMs is precisely an example of the rush to generalising effects. Instead, as the CSCW literature recommends, careful analysis of actual and potential use, primarily through qualitative methods, gives us more nuanced results. We aim to do this in what follows.

## 3 Methodology

Our paper emerged from a general interest in the possibilities and challenges inherent in the latest iterations of machine learning algorithms, generative AI, and the recognition that we have, yet very little data concerning how practitioners and users in relevant domains see the prospects. To this end, we used opportunity sampling to identify respondents in the financial services industry and more specifically working in Fintech, to talk to them about their perceptions. Data was gathered from semi-structured interviews with personnel from 6 Fintech companies based in London, Stockholm and Copenhagen. The interviews specifically focused on adoption levels of LLMs in general, and ChatGPT in particular, in the Fintech industry, especially in the light of the current hype surrounding them, as rehearsed above. Data analysis was done initially by the first author and themes shared with the second author after a first pass on the data. Together, the various themes identified were grouped into four main themes, as detailed below. Our coding process was an informal one but generally consistent with the inductive/abductive reasoning process associated with the likes of grounded theory, thematic analysis, and so on. Given, however, that all interview data was collected by the first author, and that she also produced the first analytic results, we adopted the general inductive approach suggested by Thomas (2003; 2006). Thomas argues, consistently with the general approach to be found in other inductive or abductive approaches, that the purposes of an inductive approach are "to (a) condense raw textual data into a brief, summary format; (b) establish clear links between the evaluation or research objectives and the summary findings derived from the raw data; and (c) develop a framework of the underlying structure of experiences or processes that are evident in the raw data." It constitutes "a simple, straightforward approach for deriving findings …". Given that our study is small scale, exploratory, and involves no theory building, such a simple analytic method seemed entirely appropriate. We regard the Fintech industry as a 'perspicuous' setting for the investigation of the opportunities for, and barriers to, the uptake of new technology such as the LLM, throwing light onto real world issues that may or may not reflect academic concerns.

### 3.1 Data collection

We adopted an interviewing strategy for practical reasons. Access to Fintech sites is difficult, to say the least, especially given that much of what was under discussion is prospective. Our approach to the interviews was to adopt a semi-structured strategy, where some guiding questions decided on beforehand were used to stimulate discussion (Bryman et al. 1988, Fontana and Frey, 1994; Holstein and Gubrium 1997; Martin and Turner 1986; Alvesson 2003). We nevertheless allowed our respondents to go wherever they deemed fit during the interviews. We were keen to glean information about the way in which individual institutions and their representatives either used ChatGPT, or saw potential future uses for it (Bansal et al. 2012; Flick,2014). There was, on our part, a desire to focus on specific adoption models or approaches– for each of the six participating Fintech institutions and, as such, the interviews can be thought of as 'focused' (Brinkmann, 2013).

We set out, then, to represent views about daily usage of LLMs in general in financial institutions, while at the same time also highlighting the existing nuances and tensions in the Fintech industry's relationship with LLM technology as an agency for change in existing financial practices. In effect, given the research question we describe above, nd the very topical element of our study, the interview approach was the only 'fitting' alternative available to us. Not least, much of the planning being done around the possibilities will remain, perforce, 'commercial and in confidence', limiting the detail we can provide.

We recruited Fintech informants through our social network. Given the fact that the financial sector is quite regulated – compared to other sectors- it was quicker to use our private networks to gain access. Snowball sampling also helped recruit friends of friends for the interviews. The study informants were chosen with the caveat that they had previously adopted, or intended shortly to adopt, LLMs (and ChatGPT in particular) in their day-to-day operations. All participants lived and worked either in Denmark or the United Kingdom. Table 1 below shows participants details (suitably anonymized). Participants were promised that no identifying detail would be used, (and agreed before the interviews) although data about gender and the type of organization were retained. It was essential to us that our participants be selected for good relevant reasons, hence we focused on interviewing people who possessed a good round knowledge of LLM applications within their business operations in their institution.

Table 1 - Empirical data gathering details

| Respondent Name (Anonymised) | Gender | Organisation Type | Organisation base | Interview length |
|---|---|---|---|---|
| P | Male | Bank | Denmark | 44:18 |
| M | Male | Wealth Management Company | Denmark | 39:03 |
| O | Male | Venture Capital | UK and Sweden | 31:19 |
| P2 | Male | Startup | Denmark | 37:26 |
| MN | Male | Startup | Denmark | 62:53 |
| N | Male | Startup | Denmark | 36:01 |

The semi-structured interviews with each participant (each one representing a different Fintech institution) individually lasted between 36 and 63 minutes. All bar one of the interviews were conducted online using the Zoom platform and 3 of the 6 interviewees allowed us to both record the image and voice during our conversations. The Interviews focused on 3 main areas. The first area investigated how LLMs had been previously used in the said financial institution (e.g., who uses it, when and how). The second area of focus was to do with the vision for expanding ChatGPT usage (if applicable) within other areas of operation -in the business- and the potential legal challenges in doing so. We asked for a detailed account of how they planned on using ChatGPT as a tool in the immediate future. The final area focused on examining whether current regulations such as the existing General Data Protection Regulation (GDPR) in the EU were fit for purpose when it comes to regulating LLMs.

We further wanted to gain an understanding of what and how the main actors in the Fintech industry anticipate their future relationship with tools such as ChatGPT would be while remaining compliant within their business operations/model and regarding wider compliance issues as they are currently understood. As part of our interviews, we also asked if the development of such tools as ChatGPT was in line with current green transition efforts. In asking this question, we were trying to get a sense of priorities in the adoption and development of LLMs versus the sustainability issue/agenda in the financial sector.

### 3.2 Data Analysis

In keeping with the general inductive analysis approach we adopted, the first author first transcribed and translated the interview data into English. Subsequently, and through an iterative cycle, the two authors worked together on the data, identifying areas of similarity and hence grouping them into various categories. This involved discussing and refining our decisions about how best to conceptualize what we had discovered, what we had omitted from our first attempt at analysis, what redundancy there was in our thinking, and so on. After three such rounds, we established five main categories. We discuss them below.

## 4  Findings

### 4.1 Caution regarding adoption of LLMs

The Fintech industry recognizes the potential of ChatGPT, and LLMs in general, but is very much aware of the potential for error or bias. For the most part, this is not because of concerns about customer relations, the possibility of legal challenges or concerns over privacy, because, as one of our informants put it, "customers' data will still be held in private networks with no access to the outside world." Nevertheless, the adoption process is a guarded one. Some see the adoption process as involving, in the first instance, the management of very routine tasks. As Nicholai put it, "…*I use ChatGPT as an assistant, to help me write documents; I let it do that sort of thing that I am not an expert in … but it could never do what I do*…"

Other informants added that ChatGPT was 'a great tool' for tasks such as basic programming, process documentation, inbound customer query resolution (customer service efficiency), profiling companies to invest in, and customer spending analysis. Other, more interesting, possibilities, have to do with basic commercial customer service tasks. As one respondent (P) put it, "*We could use it for tracking customer movements and consumption patterns— are they using mobile applications? Are they moving home? Are they travelling? … Are they moving in with somebody else who's also a customer?* …"

P also suggested, "… *what we have done until now with it mainly is look at the patterns of the customers behaviour and the financial behaviours, meaning customers saving up. Are they spending more money, are they moving and logging more actively into the mobile bank? Are they visiting*?". He went on to say, "…*we used that (LLM) to figure out what is the pattern for a customer who is about to buy his first home for example. What does that look like? Are they saving up for a lot longer than 6 months, one year prior to buying their first hom*e?"

There is, then, a general agreement that ChatGPT should be used initially for mundane tasks such as process documentation, financial legal document collation or customer service data analysis, rather than anything such as full-blown financial advice, which may entail regulatory or legal risk. The latter possibilities are viewed as more problematic. Morten, who works for a company that provides wealth management service to niche clients, said, "… *I will never let ChatGPT do decision trees for me, for instance, without me checking it … that would just be plain stupid…*"

Reservations, as we have seen, were expressed, however, about tasks that are subject to legal challenge, such as financial advice. This is very much an issue of accountability and, more particularly, regulation. This is echoed in a further comment,

"*If 'it' (ChatGPT) can help me find the decision quicker and make my decision tree quicker, then I will use it, but to take the decision itself … well, we have…It's like self-driving cars we can't afford mistakes with an investment or branding, and if you don't know what happened? … It's safety first so there is a sense that we need the human, you know…The human must take the decision*"

This respondent shows an acute awareness of the consequences of potential error and remains confident that human experience and skill must be involved in the final decision.

**4.2 Existing regulations are not fit for purpose**

Regulation was a major issue for all our respondents. 5 out of 6 of our respondents mentioned that more needs to be done to regulate advances in LLM development. One respondent in Denmark went further in denouncing regulators in the US, suggesting they were perhaps too elderly and did not understand the technology well enough to be proactive with regulation. Hence:

"*You look at the American Congress, Old people, 80 years of age on average. A senator of 70 or something, they don't understand something has gone wrong and regulations come at the back of catastrophes - the financial crisis in 2008 for instance. We must regulate the banks. OK, you should have regulated 10 years before that happened, it is like chasing a tail.*"

Respondents in general were of the view that, in the financial services industry, regulation was always 'one step behind' and were always having to play catch up with technological innovation. Thus:

"*ChatGPT will go the same way as other technologies … the regulators will be trying to keep up … there will be a discussion about that and the Americans are way behind the European when it comes to regulation, so, so we should look at the EU because that's 350 million (people). This economic zone and others, such as the Islamic zone, will make a difference when they take some responsibility. The EU is doing a lot of work right now on a lot of the problems*".

Of course, what is critical for our respondents is not just the need for regulation but, just as much, an examination of why there is a need. Not least, financial institutions can be made accountable for their

decision- making processes both by regulatory agencies and by their customers (who may be both retail customers and other commercial interests). In addition, in a competitive environment, trust in their processes by customers is regarded as a paramount issue. Data itself is not seen to be the problem since, as is pointed out above, it will be stored privately, but its deployment in situations where inaccuracy could be consequential, is. Our respondents were aware of the problem of 'ground truth' and much of the caution they expressed was a result of concerns about the inaccuracy that could result when various kinds of correlative work are undertaken.

**4.3 Bespoke versions**

All 6 fintech companies welcomed the advent of ChatGPT but were keen to develop their own version to better control the data flow within it and thus "protect our customers." There is a general view that over time companies will implement ways to customize the tool to meet their own business needs, training models which use their own datasets. Some of our respondents went further and elaborated on an opportunity to build what they called "a competitor" to ChatGPT for Fintech. Martin, for instance, conceded : "... *So essentially building an open GPT4 or at least a competitor or something as close as we can get. Kind of the strong way of thinking...*"

He later added "... *what is important in terms of getting good performance in a particular field or a particular domain requires a little bit more. Some of it you can do by doing again and again* ..."

All our respondents saw ChatGPT and similar LLMs as useful tools for training bespoke models. The primary motivation for this was to better control the data flow within it, ensuring data validity and reliability. Our participants repeatedly explained that they were keen to build an in-house, secured and controlled version of ChatGPT to adhere to the high demands emanating from regulation in this industry. Regulation occurred repeatedly as a dominant theme in our discussions, both because there was an evident need to abide by such rules as existed but just as much because of the current ambiguities about what those rules were and how they might be implemented.

**4.4 Green efforts are not a priority now, when it comes to LLMs development.**

That LLMs are energy intensive, have significant impact on water supplies etc. is by now unarguable (see Annany and Crawford, 2018). Similar concerns have been expressed over the army of workers engaged in data input, largely in the global south (Crawford, 2021). Nevertheless, and despite the existence of a so-called 'green Fintech' agenda espoused by the EU (see e.g. Macchiavello) and Siri, 2022), the fintech actors we spoke to agreed that green efforts were the least important aspect on the agenda when considering LLMs and their application, although one respondent in Denmark urged us to report on the water wastage element in the development of such tools. For the most part, our respondents saw such issues as beyond their control:

Morten: "*The globe is going under ... that is so political. There is no consensus, and it (ChatGPT) cannot come up with a consensus.*" Nicholai similarly said: "*Unfortunately, I don't think it's a priority right now. I think the priority is on building infrastructure and then you know, they deal with the symptoms of that afterwards and symptoms there are being, I mean, water and electricity.*"

Others saw some positive possible benefits: "... *it could help biologists find unknown species in the future.', and 'even up until today, we mapped out 5% of all chemicals in nature. So, there's a lot more to discover. And if we(society) have a chance to do that, we need some help (from LLMs)...*"Ola.

This speaks to the difference between a set of broadly academic interests focusing on the moral and political consequences of LLMs, something that has been remarked upon in a wide range of contexts (see for instance, Punchoojit et al, 2015; Nunes Vilaza et al, 2022) and the much more pragmatic view evinced by our respondents. They made it very clear, rightly or wrongly, that they did not feel that they had any control over outcomes and hence did not regard ethical consequences in relation to energy use as central to their concerns.

**Discussion and conclusion**

It is by now a commonplace in HCI and CSCW that technologies need to be examined in relation to the attitudes, values and practices of those who will use them. Put simply, we need use cases. Although some research has taken place into possible uses of LLMs in, for instance, medical imaging (e.g. Yang et al, 2023); education (Su and Wang, 2023); radiology (Elkassem and Smith, 2023), and software requirements (Krishna et al, 2024), we can find few examples of qualitative research which examines the practitioner viewpoint in any specific domain. The need for such research is evident, given that our study demonstrates some differences between what we might term academic concerns and those of real-world practitioners. Not least, the trajectory of current thinking about LLMs arguably parallels the cycle of hype and retrenchment we have seen on more than one occasion before. Gartner, in 1995, referred to 'hype cycles' and made the point that there were stages to these cycles, involving an innovation trigger, what they refer to as 'a peak of inflated expectations' followed by a 'trough of disillusionment' and finally a slow slope of enlightenment leading to a final plateau (see also Linden and Fenn, 2006; Dedehayir and Steinert, 2016)

We have, that is, been here before. Thus, and for instance, in 1970, Marvin Minsky was quoted as saying, "In from three to eight years we will have a machine with the general intelligence of an average human being" and went on to say, "If we are lucky, they might decide to keep us as pets.".

Such fears are being rehearsed yet again in respect of the so 'called 'singularity' (see Kurtweil, 2005). Early proposals for Japanese 5th generation AI objectives, the so-called Fifth Generation Computer Systems project (FGCS), which commenced in 1982, were similar to those advanced for LLMs, and included ambitions for problem-solving and inference, knowledge-base management, and intelligent interfaces (including natural language processing).Therefore, according to Garvey (2019), the new machines would need to have the "ability to process information conversationally using everyday language" and visual media, such as imagery and video, to teach its non-expert users. They must move, that is, from numerical computations to machines that can assess the meaning of information and understand the problems to be solved. Garvey (ibid), amongst others, has reviewed the relative success and failure of this project and pointed out its substantial effect on the subsequent direction of much research, largely because of fears about economic power and the global marketplace. In a second development, associated with Feigenbaum (see Buchanan et al, 2006) the search became one for decision support systems. This entailed a narrower focus, one where problems to be resolved were to be found in bounded knowledge domains. Nevertheless, it is striking that, some 40 plus years after the initial promise of decision support systems was asserted, reservations are still being expressed. Hence, Antoniadi et al (2021) point to the continuing problem of 'explainability' in such systems when deployed in support of medical work. Problems include lack of knowledge about training data; trust and uncertainty, and the need for multidisciplinary evaluation of the knowledge generated. None of this is to deny the very real progress that has been made, but we note that typically, much as the Gartner hypothesis suggests, real-world applications take time to come to fruition.

Our purpose, then, in the above was to compare a practitioner view of the opportunities and challenges with those prevalent in the literature. Needless to mention that, for all our informants, finance and innovation was at the forefront of their concerns and LLMs were seen as a potential tool to help them achieve their goals. Nevertheless, their views can be summed up as pragmatic. Although our respondents, for instance, demonstrated an awareness of the ecological ramifications of LLMs, they did not for the most part see these issues as anything that they could or should manage. Similarly, security and privacy did not feature as a major concern, largely because none saw 'open' LLMs as being useful for their purposes for anything other than routine clerical or administrative work. When assessing other possible uses, notably those associate with customer relationship management, reliability did become an issue. This was for two distinct, but related reasons. The first, evidently, had to do with the customer. Their hesitation comes about from the desire to 'protect our customers', which reflects the acknowledgement of the risks which emanate from the quality and structure of the data itself, most notably in relation to the kinds of correlative data which might be thrown up. Hence, respondents pointed to the need for accuracy when it came to the reporting of customer purchasing behaviour and the demographic and other factors which relate to it. Similarly, they saw the need for care when it came to the use of LLMs for branding and for investment advice. The latter prospect we treated with some suspicion. Related to this is the problem of regulation. All our respondents, without exception, currently stressed the problematic relationship between new technology such as LLMs and existing regulatory frameworks, which in their current form they saw as largely inadequate. This matters because it imposes forms of accountability on institutional representatives, and they remain unwilling to countenance more sophisticated uses of LLMs until they can judge the specific ways in which they and their institutions may be made accountable.

It is clear from the above that our respondents all saw potential benefit from the advent of LLMs but at the same time, they identified some considerable risk, mainly in relation to their relationships with customers and the legal frameworks which determine what that relationship will look like. For this reason, and unlike those who have engaged in the 'hype' surrounding LLMs, our respondents were uniformly cautious.

In sum, in this paper we have offered a consideration of the practical ways in which those in the Fintech industry currently see the uses and limitations of Large Language Models. Their caution, for the reasons we have given, is understandable. The informants from the Fintech industry we have spoken to have an obligation towards their clients, and thus must think of ways to make this technology fit for purpose while remaining accountable to regulators in their daily operations. We suggest that, despite the perceived hype concerning LLMs introduced in certain industries such as advertising, the Fintech industry remains cautious and guarded in relation to the adoption of LLMs such as ChatGPT. Our empirical material with the stakeholders in said industry demonstrates that there is a potential for a customised or 'bespoke' future version of LLMs that will meet the regulatory demands they work with, but there is still a long way to go before such ambitions become reality.

**ACKNOWLEDGMENTS**

The authors wish to thank their informants for participating in this study. Without the informants' invaluable input, it would not have been possible to write this paper.